\documentclass[conference, compsoc]{IEEEtran}
\usepackage{graphicx}
\usepackage{balance}
\usepackage{subfigure}
\usepackage{epsfig}
\usepackage{wrapfig}
\usepackage{amsfonts}
\usepackage[cmex10]{amsmath}
\usepackage[cmex10]{amsmath}
\usepackage{amscd}
\usepackage{dsfont}
\usepackage{amssymb}

\begin{document}
%
% paper title
% can use linebreaks \\ within to get better formatting as desired
\title{Inferring Information from Feature Diagrams to Product Line Economic Models}

% author names and affiliations
% use a multiple column layout for up to two different
% affiliations

%\author{\IEEEauthorblockN{David Fernandez-Amoros}
%\IEEEauthorblockA{Dpto. de Lenguajes y Sistemas Informaticos\\
%    Universidad Nacional de Educacion a Distancia\\
%    Madrid, Spain\\
%    david@lsi.uned.es}
%\and
%\IEEEauthorblockN{Ruben Heradio-Gil,Jose Cerrada}
%\IEEEauthorblockA{Dpto. de Ingenieria de Software y Sistemas Informaticos\\
%    Universidad Nacional de Educacion a Distancia\\
%    Madrid, Spain\\
%    rheradio@issi.uned.es}
%}
\author{\IEEEauthorblockN{David Fernandez-Amoros\IEEEauthorrefmark{1},
Ruben Heradio Gil\IEEEauthorrefmark{2} and
Jose Cerrada Somolinos\IEEEauthorrefmark{2}}
\IEEEauthorblockA{ETS de Ingenieria Informatica, Universidad Nacional de Educacion a Distancia, Madrid, Spain}
\IEEEauthorblockA{\IEEEauthorrefmark{1}david@lsi.uned.es}
\IEEEauthorblockA{\IEEEauthorrefmark{2}(rheradio$|$jcerrada)@issi.uned.es}
}
% make the title area
\maketitle
\begin{abstract}
Existing economic models support the estimation of the costs and benefits of developing and evolving a Software Product Line (SPL) as compared to undertaking traditional software development approaches. In addition, Feature Diagrams (FDs) are a valuable tool to scope the domain of a SPL. This paper proposes an algorithm to calculate, from a FD, the following information for economic models: the total number of products of a SPL, the SPL homogeneity and the commonality of the SPL requirements. The algorithm running time belongs to the complexity class $O(f^42^c)$. In contrast to related work, the algorithm is free of dependencies on off-the-self tools and is generally specified for an abstract FD notation, that works as a pivot language for most of the available notations for feature modeling.
\end{abstract}

% For peerreview papers, this IEEEtran command inserts a page break and
% creates the second title. It will be ignored for other modes.
\IEEEpeerreviewmaketitle

\section{Introduction}
Software Product Line (SPL) practice is a widely used approach for the efficient development of whole portfolios of software products \cite{macgregor}. However, the SPL approach is not always the best economic choice for developing a family of related systems. The domain of a SPL must be carefully scoped, identifying the common and variable requirements of its products and the interdependencies between requirements. In a bad scoped domain, relevant requirements may not be implemented, and some implemented requirements may never be used, causing unnecessary complexity and both development and maintenance costs \cite{generative_programming}. To avoid these serious problems, SPL domains are usually modeled by mean of Feature Diagrams (FDs). Moreover, decision makers must be able to predict the costs and benefits of developing and evolving a SPL as compared to undertaking traditional development approaches. Thus, domain models are used in conjunction with existing economic models, such as the Structured Intuitive Model for Product Line Economics (SIMPLE)\cite{simple} and the Constructive Product Line Investment Model (COPLIMO)\cite{coplimo}, to estimate SPL costs and benefits.

A fundamental input parameter for economic models, that can be inferred from domain models, is the total number of products of a SPL. For instance, SIMPLE estimates the cost of building a SPL using equation \ref{eq:simplecost}, where: $C_{\mathrm{org}}$ expresses how much it costs for an organization to adopt the SPL approach, $C_{\mathrm{cab}}$ is the cost of developing the SPL \emph{core asset base}\footnote{According to the SPL approach, products are built from a \emph{core asset base}, a collection of artifacts that have been designed specifically for reuse across the SPL.}, $n$ is the number of products of the SPL, $C_{\mathrm{unique}}(\mathrm{product_i})$ is the cost of developing the unique parts of a product, and $C_{\mathrm{reuse}}(\mathrm{product_i})$ is the development cost of reusing core assets to build a product.
\begin{multline}
\label{eq:simplecost}
    C_{\mathrm{SPL}} = C_{\mathrm{org}} + C_{\mathrm{cab}} + \\
    \sum_{i=1}^{n} (C_{\mathrm{unique}}(\mathrm{product_i}) + C_{\mathrm{reuse}}(\mathrm{product_i}))
\end{multline}
Another interesting SIMPLE metric is \emph{homogeneity}, that provides an indication of the degree to which a SPL is homogeneous (i.e., how similar are the SPL products). Homogeneity is calculated by equation \ref{eq:homogeneity}, where: $n$ is the number of products of the SPL, $\|R_U\|$ is the number of requirements unique to one product, and $\|R_T\|$ is the total number of different requirements.
\begin{equation}
\label{eq:homogeneity}
\mathrm{Homogeneity_{SPL}} = 1-\frac{\|R_U\|}{\|R_T\|}
\end{equation}
Hence, unique SPL requirements $R_{U}$ must be identified in order to calculate homogeneity. Nevertheless, not only it is interesting the distinction between common and unique requirements, but also the relative importance of any requirement to the SPL, i.e., its \emph{commonality} \cite{benavides07}. Commonality of a requirement $R_j$ is calculated by equation \ref{eq:commonality}, where: $\|P_{R_j}\|$ is the number of products that implement the requirement and $n$ is the total number of products of the SPL.
\begin{equation}
\label{eq:commonality}
\mathrm{Commonality_{R_j}} = \frac{\|P_{R_j}\|}{n}
\end{equation}
This paper proposes a time-efficient algorithm to calculate, from a FD that scopes the domain of a SPL, the total number of products of the SPL, the SPL homogeneity and the commonality of the SPL requirements. In order to make our proposal as general as possible, the algorithm is specified for an abstract notation for FDs, named Neutral Feature Tree (NFT), that works as a pivot language for most of the available notations for feature modeling.

The remainder of this paper is structured as follows. Section \ref{sec:nft} formally defines the abstract syntax and semantics of NFT. Sections \ref{sec:calculuswithout} and \ref{sec:calculuswith} present the sketch of the algorithm. Section \ref{sec:relatedwork} compares our work to related research on the automated analysis of FDs. Finally, section \ref{sec:conclusions} summarizes the paper and outlines directions for future work. 
\section{An abstract notation for modeling SPL variability} \label{sec:nft}

Since the first FD notation was proposed by the FODA methodology in 1990 \cite{foda}, a number of extensions and alternative languages have been devised to model variability in families of related systems:

\begin{enumerate}
\item  As part of the following methods: FORM \cite{form}, FeatureRSEB \cite{feature_rseb}, Generative Programming \cite{generative_programming}, Software Product Line Engineering \cite{pohl}, PLUSS \cite{pluss}.
\item In the work of the following authors: M. Riebisch et al. \cite{riebisch}, J. van Gurp et al. \cite{van_gurp}, A. van Deursen et al. \cite{van_deursen}, H. Gomaa \cite{gomaa}.
\item As part of the following tools: Gears \cite{gears} and pure::variants \cite{pure_variants}.
\end{enumerate}

Unfortunately, this profusion of languages hinders the efficient communication among specialists and the portability of FDs between tools. In order to face this problem, P. Schobbens et al. \cite{schobbens07,heymans08,metzger07} propose the Varied Feature Diagram$^+$ (VFD$^+$) as a pivot notation for FDs. VFD$^+$ is expressively complete and most FD notations can be easily and efficiently translated into it. VFD$^+$ diagrams are single-rooted Directed Acyclic Graphs (DAGs). However, our algorithm takes advantage of FDs structured as trees. For that reason, we propose the usage of a VFD$^+$ subset, named Neutral Feature Tree (NFT), where diagrams are restricted to be trees.

In this section we formally define NFT. Concretely: section \ref{sec:anatomy} outlines the main parts of a formal language; sections \ref{sec:syntax} and \ref{sec:semantics} define the abstract syntax and semantics of NFT, respectively; and section \ref{sec:equivalence} demonstrates the equivalence between NFT and VFD$^+$. We emphasize NFT is not meant as a user language, but only as a formal ''back-end'' language used to define our algorithm in a general way.

\subsection{Anatomy of a formal language}  \label{sec:anatomy}

According to J. Greenfield et al. \cite{greenfield}, the anatomy of a formal language includes an \emph{abstract syntax}, a \emph{semantics} and one or more \emph{concrete syntaxes}.

\begin{enumerate}
  \item The abstract syntax of a language characterizes, in a abstract form, the kinds of elements that make up the language, and the rules for how those elements may be combined. All valid element combinations supported by an abstract syntax conform the \emph{syntactic domain} $\mathcal{L}$ of a language.
  \item The semantics of a language define its meaning. According to D. Harel et al. \cite{harel}, a semantic definition consists of two parts: a \emph{semantic domain} $\mathcal{S}$ and a \emph{semantic mapping} $\mathcal{M}$ from the syntactic domain to the semantic domain. That is,  $\mathcal{M}:\mathcal{L}\rightarrow\mathcal{S}$.
  \item A concrete syntax defines how the language elements appear in a concrete, human-usable form.
\end{enumerate}

Following sections define NFT abstract syntax and semantics. Most FD notations may be considered as concrete syntaxes or ``views" of NFT.

\subsection{Abstract syntax of NFT} \label{sec:syntax}

\begin{figure}[htbp]
    \begin{center}
      \includegraphics[width=0.5\textwidth]{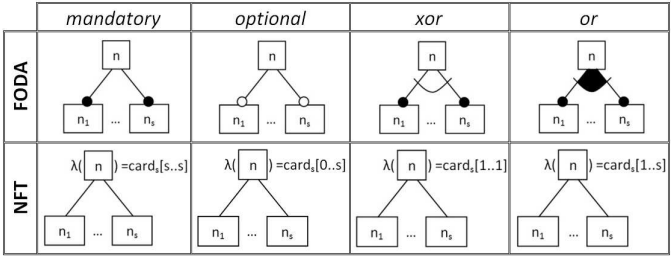}
      \caption{\emph{card} operator generalizes \emph{mandatory}, \emph{optional}, \emph{or} and \emph{xor} dependencies}\label{fig:foda}
    \end{center}
\end{figure}

\begin{figure}[htbp]
    \begin{center}
      \includegraphics[width=0.5\textwidth]{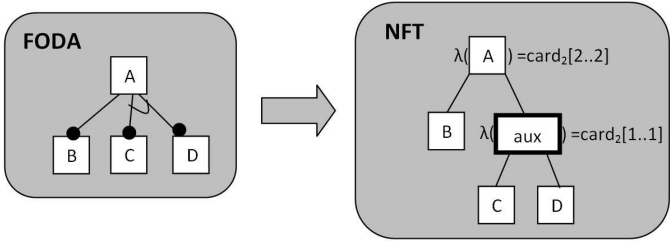}
      \caption{Different types of dependencies between a node and its children can be expressed in NFT by introducing auxiliary nodes}\label{fig:multiple}
    \end{center}
\end{figure}

A NFT diagram $d \in \mathcal{L}_{\mathrm{NFT}}$ is a tuple $(N, \Sigma, r, DE, \lambda, \phi)$, where:
\begin{enumerate}
    \item $N$ is the set of nodes of $d$, among $r$ is the root. Nodes are meant to represent \emph{features}. The idea of feature is of widespread usage in domain engineering and it has been defined as a ``distinguishable characteristic of a concept (e.g., system, component and so on) that is relevant to some stakeholder of the concept" \cite{generative_programming}.
    \item $\Sigma \subset N$ is the set of \emph{terminal nodes} (i.e., the leaves of $d$).
    \item $DE \subseteq N \times N$ is the set of \emph{decomposition edges}; $(n_1, n_2) \in DE$ is alternatively denoted $n_1 \rightarrow n_2$. If $n_1 \rightarrow n_2, n_1$ is the \emph{parent} of $n_2$, and $n_2$ is a \emph{child} of $n_1$.
    \item $\lambda : (N - \Sigma) \rightarrow \mathrm{card}$ labels each non-leaf node $n$ with a $\mathrm{card}$ boolean operator. If $n$ has children $n_1, ..., n_s$, $\mathrm{card}_s[i..j](n_1, ..., n_s)$ evaluates to \verb"true" if at least $i$ and at most $j$ of the $s$ children of $n$ evaluate to \verb"true". Regarding the $\mathrm{card}$ operator, the  following points should be taken into account\footnote{The same considerations are valid for VFD$^+$.}:
        \begin{enumerate}
          \item whereas many FD notations distinguish between \emph{mandatory}, \emph{optional}, \emph{or} and \emph{xor} dependencies, $\mathrm{card}$ operator generalizes these categories. For instance, figure \ref{fig:foda} depicts equivalences between the feature notation proposed by K. Czarnecki et al. \cite{generative_programming} and NFT.
          \item whereas, in many FD notations, children nodes may have different types of dependencies on their parent, in NFT all children must have the same type of dependency. This apparent limitation can be easily overcome by introducing auxiliary nodes. For instance, figure \ref{fig:multiple} depicts the equivalence between a feature model and a NFT diagram. Node A has three children and two types of dependencies: $A \rightarrow B$ is \emph{mandatory} and ($A \rightarrow C$, $A \rightarrow D$) is a \emph{xor}-group. In the NFT diagram, the different types of dependencies are modeled by introducing the auxiliary node $\mathrm{aux}$.
        \end{enumerate}

   \item $\phi$\footnote{also named cross-tree constraints \cite{benavides07}.} are additional textual constraints written in propositional logic over any type of node ($\phi \in \mathbb{B}(N$)).

\end{enumerate}

Additionally, $d$ must satisfy the following constraints:

\begin{enumerate}
  \item Only $r$ has no parent: $\forall n \in N \cdot (\exists n' \in N \cdot n' \rightarrow n) \Leftrightarrow n \neq r$.
  \item $d$ is a tree. Therefore,
  \begin{enumerate}
    \item a node may have at most one parent:

     $\forall n \in N \cdot (\exists n', n'' \in N \cdot ((n' \rightarrow n)\wedge(n'' \rightarrow n) \Rightarrow n'=n''))$
    \item DE is acyclic: $\nexists n_1, n_2 \ldots, n_k \in N \cdot n_1 \rightarrow n_2 \rightarrow \ldots \rightarrow n_k \rightarrow n_1$.

  \end{enumerate}
  \item $\mathrm{card}$ operators are of adequate arities:

  $\forall n \in N \cdot (\exists n' \in N \cdot n \rightarrow n') \Rightarrow (\lambda(n) = \mathrm{card}_s) \wedge (s=\|\{(n, n')|(n, n') \in DE\}\|)$
\end{enumerate}

\subsection{Semantics of NFT} \label{sec:semantics}

Feature diagrams are meant to represent sets of products, and each product is seen as a combination of terminal features. Hence, the semantic domain of NFT is $\mathcal{P}(\mathcal{P}(\Sigma))$, i.e., a set of sets of terminal nodes. The semantic mapping of NFT ($\mathcal{M}_{\mathrm{NFT}}: \mathcal{L}_{\mathrm{NFT}} \rightarrow \mathcal{P}(\mathcal{P}(\Sigma))$) assigns a SPL to every feature diagram $d$, according to the next definitions:

\begin{enumerate}
  \item A \emph{configuration} is a set of features, that is, any element of $\mathcal{P}(N)$. A configuration $c$ is valid for a $d \in \mathcal{L}_{\mathrm{NFT}}$, iff:
      \begin{enumerate}
        \item The root is in $c$ ($r \in c$).
        \item The boolean value associated to the root is \verb"true". Given a configuration, any node of a diagram has associated a boolean value according to the following rules:
            \begin{enumerate}
              \item A terminal node $t \in \Sigma$ evaluates to \verb"true" if it is included in the configuration ($t \in c$), else evaluates to \verb"false".
              \item A non-terminal node $n \in (N-\Sigma)$ is labeled with a $\mathrm{card}$ operator. If $n$ has children $n_1, ..., n_s$, $\mathrm{card}_s[i..j](n_1, ..., n_s)$ evaluates to \verb"true" if at least $i$ and at most $j$ of the $s$ children of $n$ evaluate to \verb"true".
            \end{enumerate}
        \item The configuration must satisfy all textual constraints $\phi$.
        \item If a non-root node is in the configuration, its parent must be too.
      \end{enumerate}
  \item A \emph{product} $p$, named by a valid configuration $c$, is the set of terminal features of $c$: $p = c \cap \Sigma$.
  \item The SPL represented by $d \in \mathcal{L}_{\mathrm{NFT}}$ consists of the products named by its valid configurations ($\mathrm{SPL} \in \mathcal{P}(\mathcal{P}(\Sigma$))).
\end{enumerate}

\subsection{Equivalence between NFT and VFD$^+$} \label{sec:equivalence}

\begin{figure}[htbp]
    \begin{center}
      \includegraphics[width=0.5\textwidth]{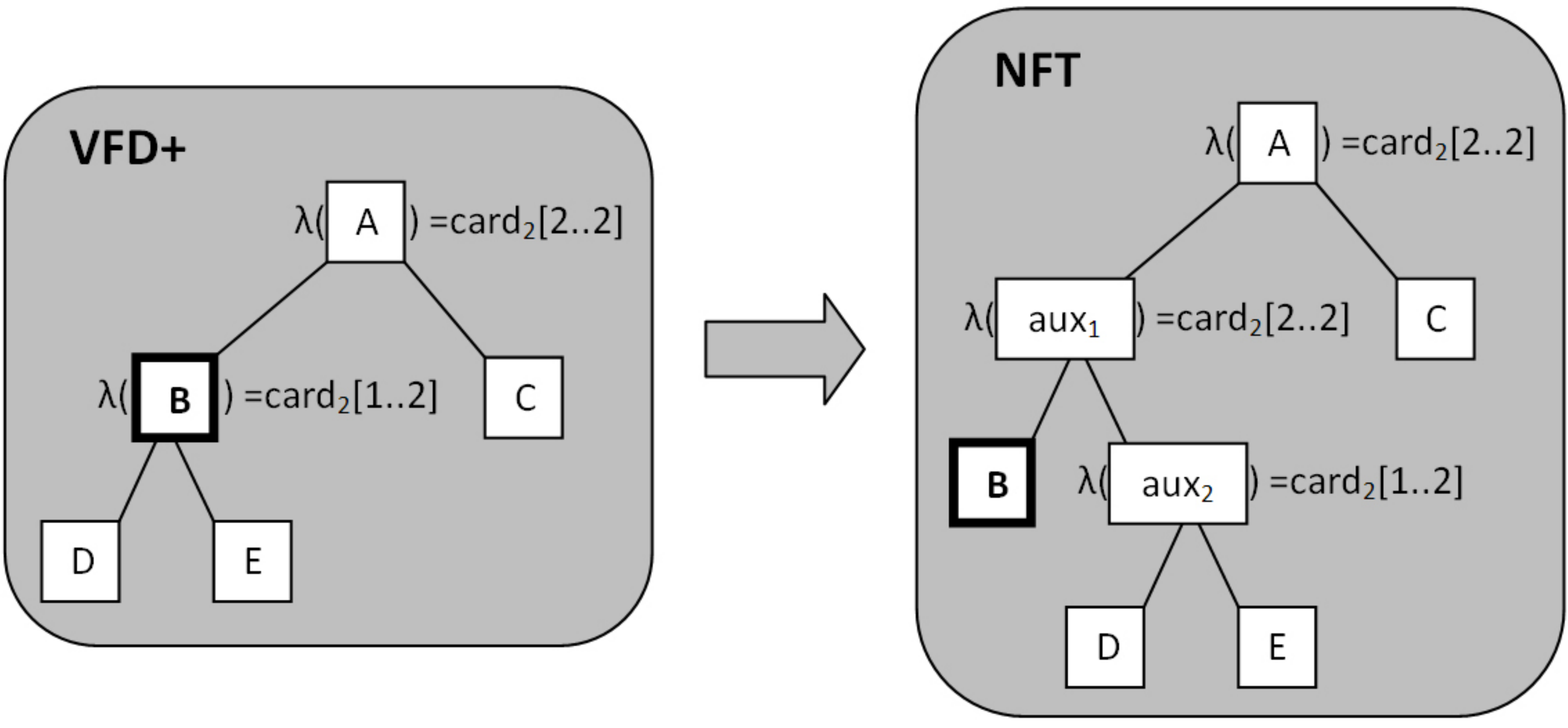}
      \caption{Any primitive non-leaf node can be converted into a leaf node by using $\mathcal{T}_{P \rightarrow \Sigma}$ }\label{fig:primitivenodes}
    \end{center}
\end{figure}

\begin{figure}[htbp]
    \begin{center}
      \includegraphics[width=0.5\textwidth]{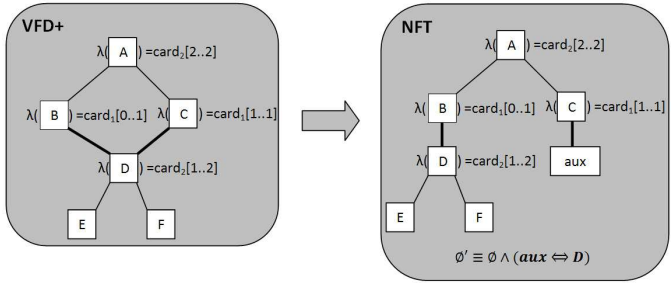}
      \caption{Any DAG can be converted into a tree by using $\mathcal{T}_{\mathrm{DAG} \rightarrow \mathrm{tree}}$}\label{fig:dag}
    \end{center}
\end{figure}

NFT differentiates from VFD$^+$ in the following points:
\begin{enumerate}
  \item \textbf{Terminal nodes vs. primitive nodes}. As noted by some authors \cite{batory05}, there is currently no agreement on the following question: are all features equally relevant to define the set of possible products that a feature diagram stands for? In VFD$^+$, P. Schobbens et al. have adopted a neutral formalization: the modeler is responsible for specifying which nodes represent features that will influence the final product (the primitive nodes $P$) and which nodes are just used for decomposition ($N - P$). P. Schobbens points that primitive nodes are not necessarily equivalent to leaves, though it is the most common case. However, a primitive node $p \in P$, labeled with $\mathrm{card}_s[i..j](n_1, ..., n_s)$, can always become a leaf ($p \in \Sigma$) according to the following transformation $\mathcal{T}_{P \rightarrow \Sigma}$:
      \begin{enumerate}
        \item $p$ is substituted by an auxiliary node $\mathrm{aux_1}$.
        \item the children of $\mathrm{aux_1}$ are $p$ and a new auxiliary node $\mathrm{aux_2}$.
        \item $\mathrm{aux_1}$ is labeled with $\mathrm{card}_2[2..2](p, \mathrm{aux_2})$.
        \item $p$ becomes a leaf. $\mathrm{aux_2}$'s children are the former children of $p$.
        \item $\mathrm{aux_2}$ is labeled with the former $\mathrm{card}_s[i..j](n_1, ..., n_s)$ of $p$.
      \end{enumerate}
      Figure \ref{fig:primitivenodes} depicts the conversion of a primitive non-leaf node $B$ into a leaf node.

  \item \textbf{DAGs vs. trees}. Whereas diagrams are trees in NFT, in VFD$^+$ are DAGs. Therefore, a node $n$ with $s$ parents $(n_1, ..., n_s)$ can be translated into a node $n$ with one parent $n_1$ according to the following  transformation $\mathcal{T}_{\mathrm{DAG} \rightarrow \mathrm{tree}}$:
      \begin{enumerate}
        \item $s-1$ auxiliary nodes $\mathrm{aux_2}, ..., \mathrm{aux_s}$ are added to the diagram.
        \item edges $n_2 \rightarrow n, ..., n_s \rightarrow n$ are replaced by new edges $n_2 \rightarrow \mathrm{aux_2}, ..., n_s \rightarrow \mathrm{aux_s}$.
        \item D. Batory \cite{batory05} demonstrated how to translate any edge $a \rightarrow b$ into a propositional logic formula $\phi_{a, b}$. Using Batory's equivalences, implicit edges $\mathrm{aux_2} \rightarrow n, ..., \mathrm{aux_s} \rightarrow n$ are converted into textual constraints $\phi_{\mathrm{aux_2}, n} ... \phi_{\mathrm{aux_s}, n}$ and are added to $\phi$ ($\phi' \equiv \phi \wedge \phi_{\mathrm{aux_2}, n} \wedge ... \wedge \phi_{\mathrm{aux_s}, n}$).
      \end{enumerate}
      Figure \ref{fig:dag} depicts the conversion of a node $D$ with two parents $B$ and $C$ into a node with a single parent.
\end{enumerate}

In order to identify when a transformation on a diagram keeps (1) the diagram semantics and (2) the diagram structure, P. Schobbens \cite{schobbens07} proposes the following definition of \emph{graphical embedding}: ``a translation $\mathcal{T}: \mathcal{L} \rightarrow \mathcal{L}'$ that preserves the semantics of $\mathcal{L}$ and is node-controlled, i.e., $\mathcal{T}$ is expressed as a set of rules of the form $d \rightarrow d'$, where $d$ is a diagram containing a defined node or edge $n$, and all possible connections with this node or edge. Its translation $d'$ is a subgraph in $\mathcal{L}'$, plus how the existing relations should be connected to nodes of this new subgraph". According to this definition, $\mathcal{T}_{P \rightarrow \Sigma}$ and $\mathcal{T}_{\mathrm{DAG} \rightarrow \mathrm{tree}}$ are graphical embeddings that guarantee the equivalency between NFT and VFD$^+$.

\section{Calculating the products in a NFT diagram without textual constraints} \label{sec:calculuswithout}

This section presents how to calculate the total number of products of a SPL modeled by a NFT diagram without considering textual constraints.

The number of products of a node $n$ is denoted as $P(n)$. Thus, the total number of products represented by a NFT diagram is $P(r)$, where $r$ is the root. For a leaf node $l$, $P(l)=1$. Table \ref{tab:noconstraints} includes equations to calculate $P(n)$ for a non-leaf node $n$ that has $s$ children $n_i$ of type \emph{mandatory} (i.e., $n$ is labeled with $\mathrm{card}_s[s..s]$), \emph{optional} ($\mathrm{card}_s[0..s]$), \emph{xor} ($\mathrm{card}_s[1..1]$) and \emph{or} ($\mathrm{card}_s[0..s]$). Hence, time-complexity for calculating $P(n)$ in these cases is $O(s)$. Therefore, time-complexity for computing $P(r)$ is linear on the diagram number of nodes, i.e., $O(N)$.
\small
\begin{table}[htbp]
    \label{tab:noconstraints}
    \begin{center}
    \renewcommand{\arraystretch}{1.5}
    \begin{tabular}{r | l}
    type of relationship & formula \\ \hline
    \emph{mandatory} ($\mathrm{card}_s[s..s]$) & $P(n) = \prod_{i=1}^{s}{P(n_i)}$ \\
    \emph{optional} ($\mathrm{card}_s[0..s]$) & $P(n) = \prod_{i=1}^{s}{(P(n_i)+1)}$ \\
    \emph{or} ($\mathrm{card}_s[1..s]$) & $P(n) = ( \prod_{i=1}^{s}{(P(n_i)+1)} ) - 1$ \\
    \emph{xor} ($\mathrm{card}_s[1..1]$) & $P(n) = \sum_{i=1}^{s}{P(n_i)}$
    \end{tabular}
    \end{center}
    \caption{Number of products for \emph{mandatory}, \emph{optional}, \emph{or} and \emph{xor} relationships}
\end{table}
\normalsize
In general, when a node $n$ has $s$ children and is labeled with $\mathrm{card}_s[low..high]$, $P(n)$ is calculated by equation \ref{eq:generalnoconstraint}, where $S_k$ is the number of products choosing any combination of $k$ children from $s$. For the sake of clarity, let us denote $P(n_1), P(n_2),\ldots P(n_s)$ as $p_1, p_2,\ldots, p_s$. In a straightforward approach, $S_k$ can be calculated by summing the number of products of all possible k-combinations (see equation \ref{eq:generalbruto}). Unfortunately, this calculation has exponential time-complexity.
\begin{equation}
\label{eq:generalnoconstraint}
P(n) = \sum_{k=\mathrm{low}}^{\mathrm{high}}{S_k}
\end{equation}
\begin{equation}
\label{eq:generalbruto}
S_k = \sum_{1 \leq i_1 < i_2 < i_3 \ldots < i_k \leq s}{p_{i_1} p_{i_2} \ldots p_{i_k}}
\end{equation}
A better complexity can be reached by using recurrent equations. The base case is $S_0 = 1$. According to equation \ref{eq:generalbruto}, $S_1 = \sum_{i=1}^{s}{p_i}$. Calculating $S_2$, the number of products for combinations of 2 siblings that include $n_1$ is $p_1p_2 + p_1p_3... + p_1p_s = p_1(p_2+p_3+...+p_s) = p_1(S_1-p_1)$. Similarly, the number of products of 2-combinations that include $n_2$ is $p_2(S_1-p_2)$. Adding up every 2-combinations, we get $\sum_{i=1}^{s}{p_i(S_1 - p_i)}$. However, in the sum each term $p_i p_j$ is being accounted for twice; once in the round for $i$ and another in the round for $j$. Thus, removing the redundant calculations:
\small
\begin{align*}
 S_2 & = \frac{1}{2}\sum_{i=1}^{s}{p_i(S_1 - p_i)}\\
     & = \frac{1}{2}(S_1\sum_{i=1}^{s}{p_i} - \sum_{i=1}^{s} p_{i}^{2})\\
     & = \frac{1}{2}(S_{1}^2 - \sum_{i=1}^{s} p_{i}^{2})\\
\end{align*}
\normalsize
Calculating $S_3$, the number of products for combinations of 3 siblings that include $n_1$ is $p_1$ multiplied by the number of products for 2-combinations that do not contain $n_1$, i.e., $p_1(S_2 - p_1(S_1 - p_1))$. Adding up every 3-combinations, we get:
\small
\[ \sum_{i=1}^{s}{p_i (S_2 - p_i(S_1 - p_i))} = S_2 S_1 - S_{1} \sum_{i=1}^{s}{p_{i}^2} + \sum_{i=1}^{s}{p_{i}^3} \]
\normalsize
This time, every triple $p_i p_j p_k$ is being accounted for three times. Hence, removing the redundant computations:
\small
\[ S_3 = \frac{1}{3}\left(S_2 S_1 - S_{1} \sum_{i=1}^{s}{p_{i}^2} + \sum_{i=1}^{s}{p_{i}^3}\right) \]
\normalsize
Our reasoning leads to the general equation \ref{eq:generalwithout}, that has a much better time-complexity $O(ks)$. Combining equations \ref{eq:generalnoconstraint} and \ref{eq:generalwithout}, we conclude that the total number of products of a SPL represented by a NFT diagram can be calculated, without considering textual constraints, in quadratic time, i.e., $O(N^2)$; what constitutes a considerable improvement from exponential to polynomical computational complexity.
\small
\begin{align}
\label{eq:generalwithout}
S_0 &= 1 \notag\\
S_k &= \frac{1}{k} \sum_{i=0}^{k-1}{((-1)^i S_{k-i-1}\sum_{j=1}^{s}{p_j^{i+1}})}\: \mbox{for }\: 1 \leq k \leq s
\end{align}
\normalsize
\begin{figure}[htbp]
    \begin{center}
      \includegraphics[width=0.35\textwidth]{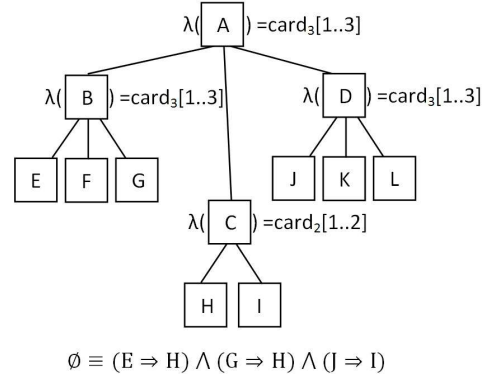}
      \caption{A sample diagram}\label{fig:davidexample}
    \end{center}
\end{figure}
Let us consider the diagram in figure \ref{fig:davidexample}. Ignoring the textual constraints in this example, it is easy to compute that nodes B and D generate 7 products each and C generates 3. Since A has \emph{or} cardinality, we could use the corresponding equation in  \ref{tab:noconstraints} and then, P(A) = (7+1)(3+1)(7+1)-1 = 255. As way of example, we will compute the same amount using equation \ref{eq:generalwithout}. We will begin computing the powers of the number of products from A's children and their sum:
\small
\begin{center}
\begin{tabular}{l|l|l|l||l|}
power & B   & C  & D   & sum\\
\hline
1     & 7   & 3  & 7   & 17\\
2     & 49  & 9  & 49  & 107\\
3     & 343 & 27 & 343 & 713\\
\end{tabular}
\end{center}
\normalsize
Now, $S_0 = 1$ by definition, $S_1 = 17$, as it is the sum of children's products, $S_2 = 1/2 (17\cdot17 - 1 \cdot 107) = 91$,  following the general formula \ref{eq:generalnoconstraint} and $S_3 = 1/3(91\cdot 17 - 17\cdot 107 + 1 \cdot 713) = 147$. Adding up $S_1$, $S_2$ and $S_3$, we get again 255.

We will now tackle another question that may be skipped in a first reading but which will of interest in the next section. Suppose we have a node N, with n children whose number of products are respectively $p_1$, $p_2$, \ldots, $p_n$. Suppose we have computed already P(N) using equation \ref{eq:generalwithout}. This calculation would provide us with vector S. What would happen if we should add a new child with $p_{n+1}$ products? We may compute a new vector S' using the general equation, but it is possible to derive $S'_i$ from $S_i$ directly, for any suitable i. Obviously, $S'_i$ will contain all the possibilities in $S_i$, since all of them are valid combinations of i children of N. These are the combinations in $S'_i$ which do not include the new node. The combinations including the new child amount to $p_{n+1}\cdot S_{i-1}$. So, $S'_i = S_i + p_{n+1}\cdot S_{i-1}$.

What we really want to do is exactly the opposite, that is, having computed S', eliminate a child and compute the vector S.
\small
\begin{align}
\label{eq:eliminatenode}
S_0 &= 1 \notag\\
S_i &= S'_i - p_{n+1} \cdot S_{i-1}
\end{align}
\normalsize
We already now S' and $S_0 = 1$ by definition. Now we can iteratively compute $S_1$, $S_2$ and so on\ldots using equation \ref{eq:eliminatenode}. Going back to our previous example, say we want to eliminate node C. Now $S_0 = 1$ by definition, $S_1 = 17 - 3\cdot 1 = 14$, $S_2 = 91 - 3\cdot 14 =  49$ and $S_3 = 147 - 3\cdot 49 = 0$ (as expected, since there are only two siblings left).

\section{Computing the number of products, commonality and homogeneity with textual constraints} \label{sec:calculuswith}
Usual SPL conceptualizations allow two types of constraints: \emph{require} and \emph{exclude}. We shall not restrain the constraints to anything other than standard propositional logic formulae.

If the constraint C is in normal conjunctive form, that is, $C \equiv C_1 \wedge C_2 \wedge \ldots \wedge C_m$, such that $C_j$ is a disjunction of literals, let $D_j \equiv \neg C_j$. Then, $D_j$ is a conjunction of literals.

Let $P(n, C)$ be the number of products in a SPL with root $n$, satisfying the constraint $C$. This function possesses two interesting properties which we will use to our advantage:
\small
\[
\begin{array}{lll}
P(n,C) & = & P(n, \mathrm{true}) - P(n,\neg C)\\
P(n, D_1 \vee D_2) & = & P(n, D_1) + P(n, D_2) \\
		   &   & - P(n, D_1 \wedge D_2)\\
\end{array}
\]
\normalsize
It is easy then to prove that,
\small
\[
P(n, \bigvee_{i=1}^{m}{D_i})  =  \sum_{K \subseteq \{1..m\}\wedge K \neq \emptyset}{(-1)^{\|K\|+1} P(n, \bigwedge_{j \in K}{D_j})}
\]
\normalsize
Now,
\small
\begin{displaymath}
\begin{array}{lll}
P(n,C) & = & P(n, \mathrm{true}) - P(n,\neg C)\\
       & = & P(n, \mathrm{true}) - P(n, D_1 \vee D_2 \vee \ldots \vee D_m)\\
       & = & P(n,\mathrm{true}) - \\
       &  &  \sum_{K \subseteq \{1..m\}\wedge K \neq \emptyset}{(-1)^{\|K\|+1} P(n, \bigwedge_{j \in K}{D_j})} \\
       & = & \sum_{K \subseteq \{1..m\}}{(-1)^{\|K\|} P(n, \bigwedge_{j \in K}{D_j})}
\end{array}
\end{displaymath}
\normalsize
If we define $D^K \equiv \bigwedge_{j \in K}{D_j}$, then
\small
\[
P(n,C) = \sum_{K \subseteq \{1..m\}}{(-1)^{\|K\|} P(n, D^K)}
\]
\normalsize
This way, we have reduced our initial problem of computing the number of products in a SPL with an unrestricted constraint to several problems in which the constraint is a conjunction of literals. As an abuse of notation, we will often drop the $D$ in $P(n, D^K)$ and write simply $P(n, K)$ wherever context is clear.

Next, we will define a series of useful concepts. Informally, we will say that a node, $n$, is \emph{selected} under a particular constraint $D^K$, and we will simply denote it by Sel(n, K), iff the particular restriction plus the structure of the tree and the associated cardinalities force the feature to be present in the products. Even if $n$ does not occur in $D^K$, $n$ may be selected because some its child nodes are. Likewise, a node $n$ will be \emph{deselected} under a constraint $D^K$, denoted by Desel(n, K)  iff $n$ does not belong to any product satisfying $D^K$, be it because it is negated in $D^K$, because the cardinality required for its child nodes is impossible to achieve or because one of its children is a contradicting node.

The constraint $D^K$ being a conjunction of literals, we will represent it by two sets, namely $A_K$ for the affirmated literals and $N_K$ for the negated literals. We also define the nodes in the subtree of a node by the function $F$. If $n$ is a node with $s$ children (with $s$ possibly being zero) $n_1, n_2, \ldots n_s$, then
\small
\[
%F(n)  =  \{ n \} \cup \bigcup_{i=1}^{s}{F(n_i)}
F(n)  =  \bigcup_{i=1}^{s}{ (\{n_i\} \cup F(n_i)) }
\]
\normalsize
It is computationally expensive to determine the number of products in a subtree of a SPL, given that we iterate over all subsets of K. Thus, we will restrict the textual constraints only to those that are relevant for the particular nodes. In order to do that, we shall define $C(n)$ as the set of constraints which affect node $n$. Let M be \{1, 2, \ldots, m\}, then
\small
\[
%R(n,K) = \{ j \in K : F(n) \cap (A_j \cup N_j) \neq \emptyset \}
C(n) = (\{n\} \cup F(n))\cap(A_M \cup N_M)
\]
\normalsize
The calculations for any given node other than the root of the tree will not involve iterating over every subset of M, as there are $2^{\mid M \mid}$ of them, but only over every subset of $C(n)$.
%Where K is, as usual, a subset of the indices for the cross-tree constraints. Note that $R(n, K)$ may exclude any constraints involving $n$. That is so because the root is implicitly affirmated, it should not be negated and that applies to subtrees as well.

Before we define formally Sel and Desel, we will also introduce some  convenient, self-explaining abbreviations:
\small
\[
\begin{array}{rlrlr}
  \mathrm{Present}(n_i, K)&\equiv&\mathrm{Sel}(n_i, K)&\wedge&\neg \mathrm{\mathrm{Desel}}(n_i,K)\\
  \mathrm{Absent}(n_i, K)&\equiv&\neg \mathrm{Sel}(n_i, K)&\wedge&\mathrm{\mathrm{Desel}}(n_i,K)\\
  \mathrm{Contradicting}(n_i, K)&\equiv&\mathrm{Sel}(n_i, K)&\wedge&\mathrm{\mathrm{Desel}}(n_i,K)\\
  \mathrm{Potential}(n_i, K)&\equiv&\neg \mathrm{Sel}(n_i, K)&\wedge&\neg \mathrm{\mathrm{Desel}}(n_i,K)\\
\end{array}
\]
\normalsize
It is desirable for a node $n$ and a constraint $D^K$ to be able to classify its child nodes according to these four possibilities. Absent nodes are not going to play a very interesting role, but the rest of them will. Let $\mathrm{Ki}$ be $K \cap C(n_i)$. This is the subset of cross-tree constraints in K which are relevant to child $n_i$ of $n$.  The set of present nodes is $\mathrm{PRE}(n, K) = \{ n_i : \:  \mathrm{Present}(n_i, \mathrm{Ki}) \}$,  here $ 1 \leq i \leq s $. We need to count how many nodes there are in each category, which we will call count-pre(n, K), count-pot(n, K) and count-con(n, K) respectively the number of present, potential and contradicting nodes. The \emph{present factor}, which we will abbreviate by \emph{pre-fac} $= \prod_{n \in \mathrm{PRE}(n, K)}{P(n, K)}$. The potential factor is the cardinality of the potential subset as explained in the previous section, with low and high readjusted to account for the present nodes.  Now, let us formally define $Sel$, $Desel$ and $P(n, K)$ for a node $n$ with cardinality card[low..high]. For leaf nodes, we just consider low and high to be zero.
\[
\mbox{Sel}(n,K) \equiv n \in A_K \vee \bigvee_{i=1}^{s}{\mbox{Sel}(n_i, \mathrm{Ki})}
\]
and
\[
\begin{array}{l}
\mbox{Desel}(n, K)  \equiv   n \in N_K \vee\\
\vee \mbox{count-pre}(n, K)  +\mbox{count-pot}(n, K) < \mbox{low} \: \vee\\
\vee \mbox{count-pre}(n, K)  > \mbox{high} \vee \mbox{count-con}(n, K) > 0\\
\end{array}
\]
The amount $P(n, K)$ will be the multiplication of the present factor and the potential factor provided there are no contradicting children.

Another interesting economic metric for SPLs is the commonality of its features. To carry out this calculation for a given feature, the number of products in which the feature appears is needed. If $n$ is a feature of a SPL and $m \in F(n)$, we define $P(n, m , D^K)$ as the number of products of the SPL with root $n$ that contain the feature $m$. This amount is really $P(n, D^K \wedge m)$. Therefore, we could follow the indications in the former part of this section to carry out the calculation. However, it is convenient to visit each node in the tree just once for each K value in order to keep computational complexity manageable.

We will first compute commonality for the children $n_i$ of a particular node, $n$,  and then we will extend that computation to every other node in $F(n)$. For a child $n_i$ of n, which is not contradicting or absent under K, the number of products depends on its siblings, that is, $P(n, n_i, K)$ will be again the product of a present factor and a potential factor, only this time $n_i$ will be considered as selected. If $n_i$ is present, the node was already selected and nothing changes wrt. the computation of P(n, K), so $P(n, n_i, K) = P(n, K)$. If $n_i$ is potential, we will have to consider it as present. We will multiply pre-fac by $P(n_i, \mathrm{Ki})$ to get the new present factor and we will eliminate $n_i$ from the potential factor as explained in the last part of the previous section, using equation \ref{eq:eliminatenode} to get the new potential factor, which we will call \emph{new-pot-fac}. Note that this promotion from potential to present can only be carried out if cardinality allows it, i.e. it is not possible if count-pre = high. In that case $P(n, n_i, K) = 0$, just as if $n_i$ was an absent node.

To compute $P(n, n_j, K)$ where $n_j \in F(n_i)$ we will proceed as in the computation of $P(n, n_i, K)$ except that the role of $P(n_i, K)$ will be played by $P(n_i, n_j, K)$.

To recapitulate, if $n$ is absent or contradicting, obviously $P(n, n_i, K) = 0$, and so are $P(n, n_j, K)$ for every $n_j \in F(n_i)$. If $n_i$ is  present under K, then $P(n, n_i, K) = P(n, K)$ and $P(n, n_j, K) = P(n_i, n_j, K) * P(n, K) / P(n_i, K)$.  The most difficult case is when $n_i$ is potential wrt. K. In that case, we compute $P(n, n_i, K)$  same as usual, only extracting $n_i$ from the list of potential nodes whose cardinality is to be computed (because it will act as a present node) and multiply said cardinality by the old \emph{pre-fac} and by $P(n_i, \mathrm{Ki})$. For a $n_j \in F(n_i)$, $P(n, n_j, K) = pre-fac * P(n_i, n_j, Ki) * \mathrm{new-pot-fac}$. Next subsection presents all these ideas in pseudocode\footnote{An executable prototype of the algorithm with source code is available on \emph{http://www.issi.uned.es/ miembros/ pagpersonales/ ruben\_heradio/ rheradio\_english.html}}.
\subsection{Algorithm specification in pseudocode}
\small
\textbf{procedure} spl(n : node) \{\\
\indent // call the children recursively \\
\indent  C(n) = $\emptyset$;\\
\indent  \textbf{foreach} child $n_i$ \textbf{of} n \textbf{do} \{ \\
\indent  \indent   spl($n_i$);  C(n) = $C(n) \cup C(n_i)$;\}\\
\indent   C(n) = C(n) $\cup \{ j | \mbox{n appears in constraint \#j} \}$ \\
\indent   // iterate over all subsets of C(n)\\
\indent  \textbf{foreach} K subset of C(n) \textbf{do} \{\\
\indent \indent   Compute $A_K$ and $N_K$;\\
\indent \indent   count-pre   = count-con = count-pot = 0;\\
\indent \indent   pre-fac     = 1;\\
\indent \indent   Sel(n, K)   = Desel(n, K) = \textbf{false};\\
\indent \indent   pot-list = $\emptyset$;\\
\indent \indent  \textbf{foreach} child $n_i$ \textbf{of} n \textbf{do} \{\\
\indent \indent \indent    Ki = $C(n_i) \cap$ K;\\
\indent \indent \indent    \textbf{if} Present($n_i$, Ki) \{\\
\indent \indent \indent \indent      count-pre++; pre-fac = pre-fac*P($n_i$, Ki); \}\\
\indent \indent \indent    \textbf{else if}  Potential($n_i$, Ki) \{\\
\indent	\indent \indent \indent  count-pot++; pot-list.add(P($n_i$, Ki)); \} \\
\indent \indent \indent \textbf{else if} Contradicting($n_i$, Ki)\\
\indent	\indent \indent \indent \indent count-cont++;\\
\indent \indent \indent     \textbf{if} Sel($n_i$, Ki)\\
\indent \indent \indent \indent     Sel(n, K) = true;\\
\indent \indent   \} // foreach child\\
\\
\indent \indent   \textbf{if} n $\in A_k$\\
\indent \indent   \indent Sel(n, K) = \textbf{true};\\	
\indent \indent   \textbf{if} n $\in N_k$\\
\indent	\indent   \indent Desel(n, K);\\
\indent \indent   \textbf{else if} count-pre $>$ n.high\\
\indent \indent   \indent        Desel(n, K) = \textbf{true};\\
\indent \indent   \textbf{else if} count-pre + count-pot $<$ n.low\\
\indent	\indent  \indent       Desel(n, bigK) = \textbf{true};\\
\indent \indent	  \textbf{else if} count-con $>$ 0\\
\indent	\indent  \indent       Desel(n, bigK) = \textbf{true};\\
\\
\indent \indent   // compute P(n, K)\\
\indent \indent   \textbf{if} Present(n, K) \textbf{or} Potential(n, K) \{\\
\indent \indent   \indent  nlow  = $max$(n.low - cont-pre, 0);\\
\indent \indent   \indent  nhigh = $max$(n.high - cont-pre, 0);\\
\indent \indent   \indent  $<$pot-fac, S$>$ = cardinality(pot-list, nlow, nhigh);\\
\indent \indent   \indent  P(n,K) = pres-factor * pot-fac;\\
\indent \indent   \}\\
\indent \indent   \textbf{else} P(n, K) = 0;\\
\\
\indent  \indent Present(n, K)      = Sel(n, K) $\wedge \neg$ Desel(n, K);\\
\indent  \indent Potential(n, K)     = $\neg$ Sel(n, K) $\wedge \neg$ Desel(n, K);\\
\indent  \indent Absent(n, K)       = $\neg$ Sel(n, K) $\wedge$  Desel(n, K);\\
\indent  \indent Contradicting(n, K) = Sel(n, K) $\wedge$  Desel(n, K);\\
\indent \indent   \textbf{if} Present(n, K) \textbf{or} Potential(n, K) \{\\
\indent  \indent \indent // compute P(n, $n_i$, K) for $n_i \in F(n)$\\
\indent  \indent \indent \textbf{foreach} $n_i$ child of \textbf{n} \textbf{do} \{\\
\indent  \indent \indent \indent Ki = $C(n_i) \cap$ K;\\
\indent  \indent \indent \indent \textbf{if} Present($n_i$, Ki) \{\\
\indent  \indent \indent \indent \indent P(n, $n_i$, K) = P(n, K);\\
\indent  \indent \indent \indent \indent \textbf{foreach} $n_j \in F(n_i)$ \textbf{do}\\
\indent  \indent \indent \indent \indent \indent P(n, $n_j$, K) =\\
\indent  \indent \indent \indent \indent \indent P($n_i$, $n_j$, Ki) * P (n, K) / P($n_i$, Ki); \}\\
\indent  \indent \indent \indent \textbf{else if} Contradicting($n_i$, Ki) \{\\
\indent  \indent \indent \indent\indent       P(n, $n_i$, K) = 0;\\
\indent  \indent \indent \indent\indent       \textbf{foreach} $n_j \in F(n_i)$ \textbf{do}\\
\indent  \indent \indent \indent\indent\indent   P(n, $n_j$, K) = 0; \}\\
\indent  \indent \indent\indent\textbf{else if} Potential($n_i$, Ki) $\wedge$ count-pre $\neq$ high \{\\
\indent	 \indent \indent \indent\indent  nnlow  = $max$ \{nlow-1, 0\};\\
\indent	 \indent \indent \indent\indent nnhigh = nhigh - 1;\\
\indent  \indent \indent \indent\indent  new-pot-fac =\\
\indent  \indent \indent \indent\indent eliminate(S, $P(n_i, \mathrm{Ki})$, nlow, nhigh);\\
\indent	 \indent \indent \indent\indent P(n, $n_i$, K) = pre-fac * P($n_i$, Ki) * new-pot-fac;\\
\indent	 \indent \indent \indent\indent\textbf{foreach} $n_j \in F(n_i)$ \textbf{do}\\
\indent  \indent \indent \indent\indent\indent P(n, $n_j$, K) =\\
\indent  \indent \indent \indent\indent\indent P(n, $n_i$, K)*P($n_i$,$n_j$,Ki)/P($n_i$, Ki); \}\\
\indent\indent\indent \indent \textbf{else} \{ //Absent($n_i$, Ki)\\
\indent \indent\indent\indent \indent P(n, $n_i$, K) = 0;\\
\indent \indent \indent\indent\indent \textbf{foreach} $n_j \in F(n_i)$ \textbf{do}\\
\indent \indent \indent \indent\indent\indent  P(n, $n_j$, K) = 0; \}\\
\indent  \indent \indent  \} // foreach $n_i$ child of n\\
\\
\indent  \indent \textbf{if} n is the root node\\
\indent  \indent \indent \textbf{foreach} $n_l$ $\in$ F(n) \textbf{do}\\
\indent  \indent \indent \indent   P(n, $n_l$) = P(n, $n_l$) + $(-1)^{|K|}$ P(n, $n_l$, K);\\
\indent  \} // if Present of Potential\\
\indent \textbf{else} \{\\
\indent \indent \textbf{foreach} $n_i$ child of \textbf{n} \textbf{do} \{\\
\indent \indent \indent P(n, $n_i$, K) = 0;\\
\indent \indent \indent       \textbf{foreach} $n_j \in F(n_i)$ \textbf{do}\\
\indent \indent \indent \indent   P(n, $n_j$, K) = 0; \}\\
\indent  \}; //foreach subset of C(n)\\ \\
\textbf{Main program}\\
spl(root);\\
homogenity = 1;\\
\textbf{if} P(root) $\neq$ 0 \{\\
\indent  \textbf{foreach} node in the diagram \textbf{do} \{\\
\indent \indent     commonality(node) = P(root, node) / P(root);\\
\indent  \indent     \textbf{if} P(root, node) = 1\\
\indent  \indent \indent       homogenity = homogeneity - 1/P(root);\\
\indent \indent \indent \}\\
\indent \}\\
\textbf{else} There are no products in the spl\\
\normalsize
\subsection{An example}
Let us consider again the diagram in figure \ref{fig:davidexample}. We enumerate the constraints $E \Rightarrow H$, $G \Rightarrow H$ and $J \Rightarrow I$, as 1, 2 and 3, respectively. We will use a bottom-up approach in order to show that $K$ iterating over all subsets of the restricted constraints is rather manageable.

Leaf nodes values are trivial; they yield one product except when they are explicitly negated. B has E, F and G as children. Thus,  F(B)=\{E, F, G\}. The constraints that apply to B are 1 and 3. Hence, we will need to compute $P(B, \emptyset), P(B, \{1\}), P(B, \{3\})$ and $P(B, \{1, 3\})$.

$D^{\emptyset}$ is simply \emph{true}. We can use the simplified version of the cardinality function, therefore
\[ P(B, \emptyset) = 2^3 -1 = 7\]
The first constraint, $E \Rightarrow H$, is equivalent to $\neg E \vee H$ and, if we negate it we get $E \wedge \neg H$ which will be our $D^{\{1\}}$. In order to compute $P(B, D^{\{1\}})$, we have to compute the number of products of the child nodes, as before. Since E, F and G are leaf nodes, they still yield one product unless negated. What has changed is that now E is a present node. Therefore,
\small
\[
\begin{array}{l}
P(B, \{1\})  = \\
= P(E, \{1\})\cdot(P(F, \{1\}) + 1)\cdot (P(G, \{1\})+1) =\\
=  1\cdot(1+1) \cdot (1+1) = 4
\end{array}
\]
\normalsize
Symmetrically, $P(B, \{3\})  = 4$. Next we compute $P(B, \{1, 3\})$. This time, both E and G are present nodes, thus:
\small
\[
\begin{array}{l}
P(B, \{1, 3\})  = \\
= (P(E, \{1, 3\}))(P(F, \{1, 3\}) + 1) \cdot P(G, \{1, 3\}) = 2\\
\end{array}
\]
\normalsize
For $C$ the situation is similar, but this time the three textual constraints are applicable (although \#1 and \#3 have the same effect on $C$).
\small
\[
\begin{array}{lll}
P(C, \emptyset)   & = & 2 \times 2 - 1  = 4 - 1 = 3\\
P(C, \{1\})       & = & 1 \times 2 - 1  = 2 - 1 = 1\\
P(C, \{2\})       & = & 2 \times 1 - 1  = 2 - 1 = 1\\
P(C, \{3\})       & = & 1 \times 2 - 1  = 2 - 1 = 1\\
P(C, \{1, 2\})    & = & 1 \times 1 - 1  = 1 - 1 = 0\\
P(C, \{1, 3\})    & = & 1 \times 2 - 1  = 2 - 1 = 1\\
P(C, \{2, 3\})    & = & 1 \times 1 - 1  = 0\\
P(C, \{1, 2, 3\}) & = & 1 \times 1 - 1  = 0\\
\end{array}
\]
\normalsize
For node $D$, the only cross-tree constraint applicable is 2.
\small
\[
\begin{array}{lll}
P(D, \emptyset) & = & 2 \times 2 \times 2 - 1 = 8 - 1 = 7\\
P(D, \{2\}) & = & 1 \times 2 \times 2  = 4\\
\end{array}
\]
\normalsize
Now that we have recollected all the necessary data about B, C and D, we finish the calculation of the number of products. Lets call $Z$ the conjunction of all the cross-tree constraints. Then,
\small
\[
\begin{array}{lll}
Z & \equiv &(\neg G \vee H)\wedge (\neg J \vee I) \wedge (\neg E \vee H)\\

P(A, Z) & = & P(A, \emptyset) - P(A, \{1\}) - P(A, \{2\})\\
 & - & P(A, \{3\}) + P(A, \{1, 2\}) +  P(A, \{1, 3\})\\
 & + & P(A, \{2, 3\}) - P(A, \{1, 2, 3\})
\end{array}
\]
\normalsize
Under $D^{\emptyset}$, B, C and D are potential nodes. Any constraint involving 1 or 3  makes B present, any constraint involving 2 makes D present and any constraints involving 1 and 2, or 3 and 2 make C absent, so we may proceed.
\small
\[
\begin{array}{lll}
P(A, \emptyset)   & = & (7+1) \cdot (3+1) \cdot (7+1) - 1 = 255\\
P(A, \{1\})       & = & 4     \cdot (1+1) \cdot (7+1)     = 64\\
P(A, \{2\})       & = & (7+1) \cdot (1+1) \cdot 4 - 1     = 64\\
P(A, \{3\})       & = & 4     \cdot (1+1) \cdot (7+1)     = 64\\
P(A, \{1, 2\})    & = & 4     \cdot 1     \cdot 4         = 16\\
P(A, \{1, 3\})    & = & 2     \cdot (1+1) \cdot (7+1)     = 32\\
P(A, \{2, 3\})    & = & 4     \cdot 1     \cdot 4         = 16\\
P(A, \{1, 2, 3\}) & = & 2     \cdot 1     \cdot 4         = 8\\
\end{array}
\]
\normalsize
Therefore,
\small
\[
P(A, Z) = 255 - 64 - 64 - 64  + 16 + 32 + 16 - 8 = 119
\]
\normalsize
Now it is time to compute how many products does a certain feature appear in. We shall calculate $P(B, E, Z)$. If we start with $D^\emptyset$, all the nodes except the root are potential, so $P(B,\emptyset) = 7$. $P(A,B,\emptyset) = P(B,\emptyset)\cdot \mbox{card}[0..2](\emptyset)(\{C,D\}) = 7 \cdot 4 \cdot 8 = 224$. Then, $P(A,E,\emptyset) = P(B,E,\emptyset) \mbox{card}[0..2](\emptyset)(\{C,D\}) = 4 \cdot 32 = 128$. Under $D^{\{1\}}$, B and G become present, so $P(A, E , \{1\}) = P(A, B, \{1\}) = P(A, \{1\}) = 64$.
Due to space limitations, we summarize the final result:
\small
\[
\begin{array}{lll}
 P(A, E, Z) & = & P(A, E, \emptyset) - P(A, E, \{1\})\\
	    & - & P(A, E, \{2\}) -  P(A, E, \{3\})\\
	    & + & P(A, E, \{1, 2\}) + P(A, E, \{1, 3\})\\
	    & + & P(A, E, \{2, 3\}) - P(A, E, \{1, 2, 3\})\\
            & = & 128 - 64 - 32 - 16 + 16 + 32 + 8 - 8\\
	    & = & 48
  \end{array}
\]
\normalsize
Hence, the commonality for E would be 48/119 = 0.403, which means that this feature appears in roughly 40\% of the products of this SPL. Finally, to calculate SPL homogeneity (see equation \ref{eq:homogeneity}) we should identify the unique features, i.e., those with a commonality value of $\frac{1}{119}$. 

\section{Computational Complexity and Related work} \label{sec:relatedwork}
%Each node in the tree has to be visited once. First, the textual constraints are restricted to those affecting the subtree of a particular node. Then, each of the subsets of the restricted constraints is iterated over to compute several attributes of the node, such as the partition of the child nodes into four categories and the number of products this node and every descendant node provide under the subset of constraints. With this information, the root node can easily compute the number of products for the SPL, the commonality for each feature and the homogeneity of the system.
The restriction of the constraints relevant for a node can be computed in linear time taking the union of those of the children plus the constraints involving the node itself. The cardinalities for a node can be computed in quadratic time, as seen in section \ref{sec:calculuswithout}. Information needed for commonality is computed inside the exponential loop of all the subsets for a particular K. This takes time proportional to $n^3 2^{\| C(n) \|}$ for each node. As every node undergoes this treatment, the time complexity for the whole SPL is $O(n^42^m)$ where n is the number of nodes and m is the number of conjunctions in the conjunctive form of the textual constraints. It is a heavy computation. Even so, this algorithm achieves an undeniable improvement over previously proposed ones which computed less properties than ours and required running times exponential to the sum of both the number of nodes and the number of constraints (limited to the \emph{requires} or \emph{excludes} flavors).

%To be fair, though, it is noteworthy to say that complexity is only of relative relevance in this algorithm since the worst cases are extremely unlikely as well as overtly pathologic: Each node only iterates exponentially over the specific subsets of the restricted constraints, which will only be the whole set for the root node. Also, in a bottom-up approach, each level only depends on the one below, not on the whole subtree. The worst case is a two-level tree in which the root has many children, with a cardinality different than \emph{optional} \emph{or} or \emph{xor} (following FODA terminology) and with many constraints between the siblings.

To the best of our knowledge, available commercial tools for SPL developing, such as Gears \cite{gears} and pure::variants \cite{pure_variants}, neither implement homogeneity nor commonality. In addition, textual constraints are not considered in the calculation of the number of products.

On the other hand, there are academic proposals for calculating commonality, homogeneity and the total number of products from a FD with textual constraints. Considering that a FD is composed of a graphical part ($g \equiv d - \phi$) and a logical part ($\phi$), most research works on the automated analysis of FDs follow one of these strategies:

\begin{enumerate}
  \item Translating the graphical part into logic formulas (i.e., $\mathcal{T}_{g \rightarrow \phi'}$) and using off-the-self tools to process the formulas and the textual constraints ($\phi \wedge \phi'$). For instance:
    \begin{itemize}
      \item D. Batory \cite{batory05} proposes a translation of FDs into propositional logic. Resulted formulas are processed by off-the-shelf Logic-Truth Maintenance Systems (LTMS) and Boolean Satisfiability (SAT) solvers.
      \item D. Benavides \cite{benavides07} devises an abstract conversion of FDs into Constraint Satisfaction Problems (CSP). FaMa Tool Suite \cite{fama} adapts this abstract conversion to general CSP solvers, SAT solvers and Binary Decision Diagrams (BDD) solvers.
    \end{itemize}
     Unfortunately, as P. van den Broek et al. \cite{broek_0} have pointed, the computation of $\phi \wedge \phi'$ to calculate the total number of products has exponential time-complexity on the size of the full FD, i.e., it belongs to the complexity class $O(2^{g + \phi})$.
  \item Embedding the textual constraints into the graphical part (i.e., $\mathcal{T}_{\phi \rightarrow g'}$) and computing the number of products from the extended graphical part. For example, P. van den Broek et al. \cite{broek_0} devise an algorithm for $\mathcal{T}_{\phi \rightarrow g'}$ where the elimination of a constraint can double the size of $g$. The algorithm time-complexity is linear on the size of $g$ and exponential on the number of textual constraints,  i.e., belongs to $O(g2^{\phi})$. Compared to our work, P. van den Broek's proposal has the following limitations:
    \begin{itemize}
        \item Only supports $\mathrm{card}_s[s..s]$, $\mathrm{card}_s[0..s]$, $\mathrm{card}_s[1..s]$ and $\mathrm{card}_s[1..1]$.
        \item Textual constraints are limited to ``A requires B" (i.e., $A \Rightarrow B$) and ``A excludes B" (i.e., $A \Rightarrow \neg B$).
    \end{itemize}
  \end{enumerate}

Table \ref{tab:comparative} presents a comparative between our proposal and related work summarized in this section.
\small
\begin{table}[htbp]
\label{tab:comparative}
\begin{center}
\begin{tabular}{c | c c c }
% cabecera
& supported & computed textual & time \\
& $\mathrm{card}$ & constraints $\phi$& complexity  \\ \hline
%Gears
Gears \cite{gears} & $\mathrm{card}_s[s..s]$ & none & unknown \\
 & $\mathrm{card_s[0..s]}$ &  &  \\
 & $\mathrm{card_s[1..1]}$ &  &  \\ \hline
%pure::variants
pure::variants & any & none & unknown \\
\cite{pure_variants} & & & \\ \hline
%fama
FaMa Tool & any & requires & $O(2^{g + \phi})$ \\
Suite \cite{fama}& & excludes & \\ \hline
%broek
Broek et al. & $\mathrm{card}_s[s..s]$ & requires & $O(g 2^\phi)$ \\
\cite{broek_0} & $\mathrm{card}_s[0..s]$ & excludes & \\
 & $\mathrm{card}_s[1..s]$ & & \\
 & $\mathrm{card}_s[1..1]$ & & \\ \hline
%this paper
our & any & propositional & $O(g^4 2^\phi)$ \\
proposal & & logic & \\ \hline
\end{tabular}
\end{center}
\caption{Comparative between our proposal and related work}
\end{table}
\normalsize
\section{Conclusions and Future Work} \label{sec:conclusions}
Existing economic models support the estimation of the costs and benefits of developing and evolving a SPL as compared to undertaking traditional software development approaches. In addition, FDs are a popular and valuable tool to scope the domain of a SPL. In this paper, we have proposed an algorithm to infer, from a FD, the following parameters and metrics fundamental for economic models: the total number of products of the SPL, the SPL homogeneity and the commonality of the SPL requirements. Instead of defining our algorithm for a specific FD notation, we have used an abstract notation named NFT, that works as a pivot language for most of the available notations for feature modeling. NFT is formally defined in this paper and it should be considered as a valuable reference notation for specifying FD analysis algorithms which take advantage of the tree organization of FDs. Compared to related work, our algorithm has a general application scope, a competitive computational time-complexity, and it is free of dependencies on off-the-self logic tools, such as LTMS and SAT solvers.

\begin{figure}[htbp]
    \begin{center}
      \includegraphics[width=0.35\textwidth]{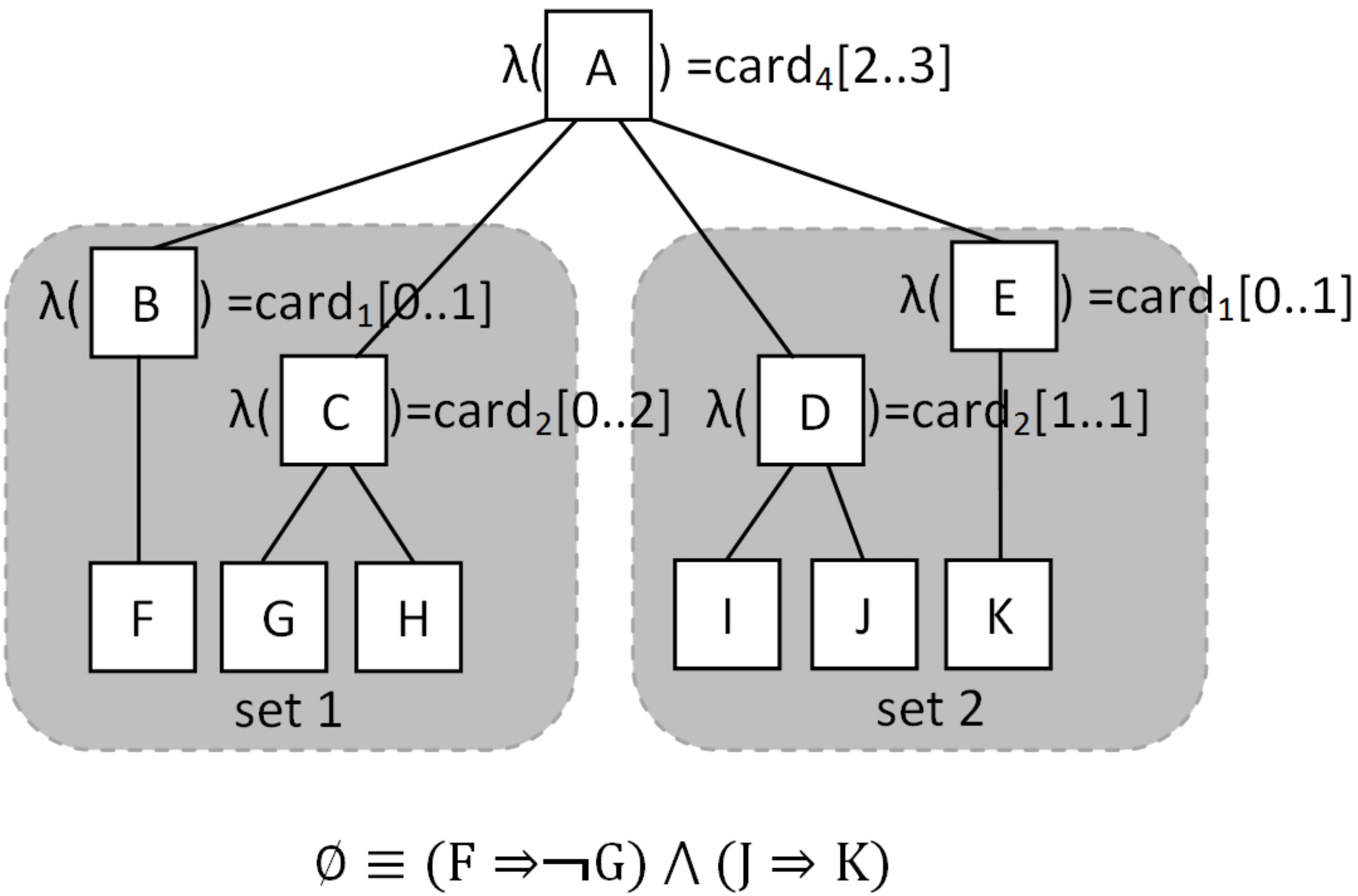}
      \caption{A 2-component diagram}\label{fig:example}
    \end{center}
\end{figure}

In the future, we plan to devise a prototype showing some improvements over the algorithm proposed in this paper. One useful and relatively simple extension would be suggesting changes to the user whenever a FD is unsatisfiable (i.e., the SPL total number of products $= 0$). It seems viable to propose minimal sets of textual constraints to be eliminated in order to achieve actual products out of the SPL. On the other hand, our efforts do not aim at improving the complexity of the algorithm, but rather its performance. For instance, in figure \ref{fig:example} the descendants of node $A$ can be divided into two forests whose textual constraints do not cross. In future work, we will try to process each of these unconnected forests separately, ignoring the textual constraints not involved in a forest and thus saving costly exponential computation.
\small
\normalsize
% This process can be of course carried out for any node, not just the root. This sort of processing is immediate for the \emph{or}, \emph{xor}, and \emph{optional} cardinalities, but we plan to focus on the more general case of the range cardinality.

\noindent\textbf{Acknowledgements.} Ruben Heradio thanks Linda Northrop for supporting
his visit to the Software Engineering Institute. We are also grateful to Felix
Bachmann, Andres Diaz Pace, Sagar Chaki and Arie Gurfinkel for their advice and insights.
 
\balance
\end{document}